\pdfoutput=1
\documentclass[final]{article}

\usepackage{times}
\usepackage{ifpdf}
\usepackage{graphics}
\usepackage{dcolumn}
\usepackage{multirow}
\usepackage{longtable}
\usepackage{amsfonts}
\usepackage{amsmath}
\usepackage{hyperref}
\input{Qcircuit.tex}

\makeatletter
\def\@normalsize{\@setsize\normalsize{12pt}\xpt\@xpt
\abovedisplayskip 10pt plus2pt minus5pt\belowdisplayskip \abovedisplayskip
\abovedisplayshortskip \z@ plus3pt\belowdisplayshortskip 6pt plus3pt
minus3pt\let\@listi\@listI}

\def\section{\@startsection {section}{1}{\z@}{10pt plus 2pt minus 2pt}
{8pt plus 2pt minus 2pt}{\centering\normalsize\sc
\edef\@svsec{\thesection.\ }}}
\def\thesection{\Roman{section}}

\def\subsection{\@startsection {subsection}{2}{\z@}{10pt plus 2pt minus 2pt}
{6pt plus 2pt minus 2pt}{\normalsize\sl
\edef\@svsec{\thesubsection.\ }}}
\def\thesubsection{\Alph{subsection}}

\long\def\@makecaption#1#2{
\vskip6pt\begin{center} #1 #2 \end{center}\par\vskip 1pt}
\def\fnum@figure{\raggedright{\footnotesize Fig. \thefigure }.%
\footnotesize}
\def\fnum@table{\footnotesize TABLE \thetable\\\footnotesize\sc}
\def\thetable{\Roman{table}}

\partopsep          \z@

\parsep             \z@

\itemsep            \z@

\labelwidth         \z@

\def\@listi{\leftmargin\leftmargini \topsep 2pt plus 1pt minus 1pt}
\let\@listI\@listi
\def\@listii{\leftmargin\leftmarginii\labelwidth\leftmarginii%
    \advance\labelwidth-\labelsep \topsep 2pt}
\def\@listiii{\leftmargin\leftmarginiii\labelwidth\leftmarginiii%
    \advance\labelwidth-\labelsep \topsep 2pt}
\def\@listiv{\leftmargin\leftmarginiv\labelwidth\leftmarginiv%
    \advance\labelwidth-\labelsep \topsep 2pt}
\def\@listv{\leftmargin\leftmarginv\labelwidth\leftmarginv%
    \advance\labelwidth-\labelsep \topsep 2pt}
\def\@listvi{\leftmargin\leftmarginvi\labelwidth\leftmarginvi%
    \advance\labelwidth-\labelsep \topsep 2pt}

\def\baselinestretch{0.97}

\makeatother

\begin{document}
%
\title{\Large\bf Decomposition of Diagonal Hermitian Quantum Gates Using Multiple-Controlled Pauli \emph{Z} Gates}
\date{}


\author{\normalsize
 \begin{tabular}[]{c c}
   \hspace{0 pt}\large Mahboobeh Houshmand & \hspace{8 pt} \large Morteza Saheb Zamani \\\\
   \hspace{0 pt} \large Mehdi Sedighi &  \hspace{8 pt} \large Mona Arabzadeh\\\\
\end{tabular}
\\
   \normalsize \textbf{\emph{Department of Computer Engineering and Information Technology}}\\
  \normalsize \textbf{\emph{Amirkabir University of Technology}}\\
    \normalsize \textbf{\emph{Tehran, Iran}}\\\\
%
  \normalsize \emph{\{houshmand,szamani,msedighi,m.arabzadeh\}@aut.ac.ir}\\
}
\maketitle
\thispagestyle{empty}

\newcounter {TCounter}
\newcounter {LCounter}
\newcounter {PCounter}
\newcounter {CCounter}
\newcounter {DCounter}
\newcounter {ECounter}

\newtheorem{theorem}[TCounter]{\textbf{Theorem}}
\newtheorem{lemma}[LCounter]{\textbf{Lemma}}
\newtheorem{proposition}[PCounter]{\textbf{Proposition}}
\newtheorem{corollary}[CCounter]{\textbf{Corollary}}
\newtheorem{definition}[DCounter]{\textbf{Definition}}
\newtheorem{example}[ECounter]{\textbf{Example}}

\newenvironment{proofs}[1][\textbf{Proof}]{\begin{trivlist}
\item[\hskip \labelsep {\bfseries #1}]}{\end{trivlist}}
\newenvironment{remark}[1][\textbf{Remark}]{\begin{trivlist}Alg:1
\item[\hskip \labelsep {\bfseries #1}]}{\end{trivlist}}

\newcommand{\qed}{\nobreak \ifvmode \relax \else
      \ifdim\lastskip<1.5em \hskip-\lastskip
      \hskip1.5em plus0em minus0.5em \fi \nobreak
      \vrule height0.75em width0.5em depth0.25em\fi}

{\small\bf Abstract---
Quantum logic decomposition refers to decomposing a given quantum gate to a set of physically implementable gates. An approach has been presented to decompose arbitrary diagonal quantum gates to a set of multiplexed-rotation gates around $z$ axis.
In this paper, a special class of diagonal quantum gates, namely diagonal Hermitian quantum gates, is considered and a new perspective to the decomposition problem with respect to decomposing these gates is presented. It is first shown that these gates can be decomposed to a set that solely consists of multiple-controlled $Z$ gates. Then a binary representation for the diagonal Hermitian gates is introduced. It is shown that the binary representations of multiple-controlled $Z$ gates form a basis for the vector space that is produced by the binary representations of all diagonal Hermitian quantum gates. Moreover, the problem of decomposing a given diagonal Hermitian gate is mapped to the problem of writing its binary representation in the specific basis mentioned above.
Moreover, CZ gate is suggested to be the two-qubit gate in the decomposition library, instead of previously used CNOT gate.
Experimental results show that the proposed approach can lead to circuits with lower costs in comparison with the previous ones.}
\\
\\
{\small\bf Keywords---
Diagonal Hermitian quantum gates, Optimization, Decomposition
}

\section {Introduction}
Quantum logic decomposition refers to decomposing a given quantum gate to a set of physically implementable gates by quantum technologies. This set of gates typically consists of CNOT and single-qubit gates, called ``basic gate" library~\cite{Barenco95} or CNOT and single-qubit rotation gates, called ``elementary gate" library~\cite{bullock2004}. Many studies have been performed on the decomposition of quantum gates. Barenco et al.~\cite{Barenco95} showed that the number of CNOT gates required to implement an arbitrary quantum gate on $n$ qubits was $O(n^3 4^n)$. In~\cite{George2001}, it was shown that applying the QR matrix decomposition in linear algebra to decompose quantum gates could lead to the same result.
The highest known lower bound on the number of CNOT gates to decompose an $n$-qubit quantum gate was reported as $\left\lceil {\frac{1}{4}(4^n  - 3n - 1)} \right\rceil$ in~\cite{shende-2004}. The improved method based on QR decomposition~\cite{vartiainen-2004} and Cosine-Sine Decomposition (CSD)~\cite{Mottonen-2004,Shende06,saeedi2010block} achieve the order of $O(4^n)$ CNOT gates.

To further reduce the number of elementary gates, the decomposition problem of more specific quantum operators, such as two-qubit gates~\cite{vidal-2004-69,Shende-1-2004,vatan-2004-69} and controlled unitary gates~\cite{song-2003-3} has been considered. Among the specific quantum gates are the diagonal quantum operators. The decomposition problem for these gates has been addressed in~\cite{Hogg1999,Bullock2003,bullock2004,Shende06}. 
In~\cite{Bullock2003} it has been shown that any two-qubit diagonal matrix can be implemented by at most five elementary gates up to a global phase. The authors of~\cite{bullock2004} prove that the lowest number of required elementary gates for decomposition of arbitrary $n$-qubit diagonal quantum gates is $2^n-1$ and present asymptotically optimal circuits for these gates which require at most $2^{n+1}-3$ elementary gates. Using the definition of quantum multiplexer gates in~\cite{Shende06}, the diagonal gates of~\cite{bullock2004} can be decomposed to a set of multiplexed-rotation gates around $z$ axis.

In this paper, a new perspective to the decomposition problem with respect to decomposing $n$-qubit diagonal Hermitian gates is presented. At first, it is shown that such gates can be decomposed to a set that solely consists  of $\mathrm{C^kZ}$ gates, $0\leq k \leq n-1$. Then a binary representation for diagonal Hermitian quantum gates is introduced and the binary representations of the set of $\mathrm{C^kZ}$ gates are mapped to a basis for vector space that is produced by these binary numbers. After that, the problem of decomposing a diagonal Hermitian quantum gate is mapped to the problem of writing these binary numbers in that specific basis.

Although CNOT is the common two-qubit gate in existing decomposition libraries, in some technologies such as ion-trap, which is one of the most promising candidates for realization of scalable quantum computers~\cite{nakahara}, CZ gate has been directly implemented~\cite{Nielsen10} and CNOT is realized by using a middle CZ gate and two rotation gates around the $y$ axis that apply on the target qubit. Therefore, a new library that contains CZ gate as a two-qubit gate can be considered. $\mathrm{C^kZ}$ gates for $k>1$ can in turn be decomposed to CZ and single-qubit gates. CZ gate is also useful in creating a parallel structure for quantum circuits using one-way quantum computation model~\cite{PhysRevLett.86,par}, as the input quantum circuits to this approach are assumed to contain CZ gates. Single-qubit gates in ion-trap technology should be constructed from rotation gates around $y$ and $x$ axes~\cite{Nielsen10}.

Diagonal gates appear in some decomposition methods such as QR \cite{vartiainen-2004} and CSD \cite{Bergholm:2004} and in the case that these diagonal gates are Hermitian, the proposed approach can be applied.

The paper is organized as follows. In the next section, some preliminaries are presented.
Section~\ref{sec:proposed} explains the proposed approach. Experimental results are presented in Section~\ref{sec:results} and finally, Section~\ref{sec:conc} concludes the paper.

\section{Preliminaries}
In this section, preliminaries about quantum states and the quantum gates used in this paper are introduced.

\subsection{Quantum States and Quantum Gates}
\label{sec:qubits}
Quantum bits or \emph{qubits} are quantum analogues of classical bits. A qubit is a unit vector in a two-dimensional Hilbert space, $\mathcal{H}_2$, for which an orthonormal basis, denoted by $\{$$\left\vert 0\right\rangle$,
 $\left\vert 1\right\rangle$$\}$, has been fixed. Unlike classical bits, qubits can be in a superposition of $\left\vert 0\right\rangle$ and $\left\vert 1\right\rangle$ like $\alpha\left\vert 0\right\rangle+\beta\left\vert 1\right\rangle$  where $\alpha$ and $\beta$ are complex numbers such that
$|\alpha|^2 + |\beta|^2 = 1$. If such a superposition is measured with respect to the basis $\{$$\left\vert 0\right\rangle$, $\left\vert 1\right\rangle$$\}$, then the classic outcome of 0 is observed with the probability of $|\alpha|^2$ and the classical result of 1 is observed with the probability of $|\beta|^2$. If 0 is obtained, the state of the system after measurement will collapse to $\left\vert 0\right\rangle$ and if 1 is obtained, it will be $\left\vert 1\right\rangle$.

Every $n$-qubit quantum gate is a linear transformation represented by a unitary matrix defined on an $n$-qubit Hilbert
space. A matrix $U$ is \emph{unitary} if $UU^{\dagger} = I$, where $U^{\dagger}$ is the conjugate transpose of the matrix $U$.
Some useful single-qubit gates are the elements of the Pauli set:
\[\sigma_0=
I=%
\begin{bmatrix}
1 & 0\\
0 & 1
\end{bmatrix},
\sigma_1=
X=%
\begin{bmatrix}
0 & 1\\
1 & 0
\end{bmatrix},
\sigma_2=
Y=%
\begin{bmatrix}
0 & -i\\
i & 0
\end{bmatrix},
\sigma_3=
Z=%
\begin{bmatrix}
1 & 0\\
0 & -1
\end{bmatrix}.\]

Another class of useful unitary gates on a single qubit are rotation operators around $x$, $y$ and $z$ axis with the angle $\alpha$ in the Bloch sphere, as shown below:

\[
R_x (\alpha )=%
\begin{bmatrix}
{{\rm cos}\frac{\alpha }{{\rm 2}}} & { - i{\rm sin}\frac{\alpha }{{\rm 2}}}\\
{- i{\rm sin}\frac{\alpha }{{\rm 2}}} & {{\rm cos}\frac{\alpha }{{\rm 2}}}
\end{bmatrix},
R_y (\alpha )=%
\begin{bmatrix}
{{\rm cos}\frac{\alpha }{{\rm 2}}} & { - {\rm sin}\frac{\alpha }{{\rm 2}}}\\
{{\rm sin}\frac{\alpha }{{\rm 2}}} & {{\rm cos}\frac{\alpha }{{\rm 2}}}
\end{bmatrix},
R_z (\alpha )=%
\begin{bmatrix}
{e^{ - i\frac{\alpha }{2}} } & 0\\
0 & {e^{ i\frac{\alpha }{2}} }
\end{bmatrix}.\]
Hadamard, $H$, and \emph{T} are two other known single-qubit gates where:
\[
H=%
\frac{1}{\sqrt{2}}\begin{bmatrix}
1 & 1\\
1 & -1
\end{bmatrix},
T=
\begin{bmatrix}
e^{i\frac{\pi }{8}} & 0\\
0 & e^{-i\frac{\pi }{8}}
\end{bmatrix}
.
\]

If $U$ is a gate that operates on a single qubit, then controlled-$U$ is a gate that operates on two qubits, i.e., control and target qubits, and \emph{U} is applied to the target qubit if the control qubit is $\left\vert 1\right\rangle$ and leaves it unchanged otherwise. For example, controlled-$Z$ (CZ) and controlled-NOT (CNOT) gates perform the $Z$ and $X$ operators respectively on the target qubit if the control qubit is $\left\vert 1\right\rangle$. Otherwise, the target qubit remains unchanged. $\mathrm{C^{k}U}$ gates have $k$ control qubits and one target. When all $k$ control qubits are in the state $|1\rangle$, gate $U$ is applied on the target qubit and no action is taken otherwise. $\mathrm{C^2NOT}$ gate is called Toffoli gate.

A quantum circuit consists of quantum gates interconnected by quantum wires carrying qubits with time flowing from left or right. The unitary matrix of the quantum circuit is evaluated by either dot product or tensor product of the unitary matrices of those quantum gates. The net effect of the gates which are applied to the same subset of qubits in series is computed by the dot product which is the same as the known matrix multiplication. The adjacent gates which act on independent subsets of qubits can be applied in parallel and their overall net effect is computed by their tensor product as defined as follows. Let $A$ be an $m\times n$ matrix and let B be a $p\times q $ matrix. Then $A \otimes B$ is an $(mp)\times(nq)$ matrix called the tensor product (Kronecker product) of $A$ and $B$ as defined below:
\[
A \otimes B = \left[ \begin{array}{l}
 a_{11} B,a_{12} B,...,a_{1n} B \\
 a_{21} B,a_{22} B,...,a_{2n} B \\
 \,\,\,\,\,\,\,\,\,\,\,\,\,\,\,\,\,\,... \\
 a_{m1} B,a_{m2} B,...,a_{mn} B \\
 \end{array} \right]
\]
where $a_{ij}$ shows the element in the $i^{th}$ column and the $j^{th}$ row of matrix $A$.
\subsection{Hermitian Quantum Gates}
\label{sec:ckz}
A matrix $\mathbb{H}$ is called Hermitian~\cite{Nielsen10} or self-adjoint if $\mathbb{H}^\dag=\mathbb{H}$. The Pauli matrices are some examples of Hermitian matrices. All eigenvalues of a Hermitian matrix are real numbers. On the other hand, the eigenvalues of a unitary matrix have modulus equal to 1 and hence the eigenvalues of a Hermitian unitary matrix are either +1 or -1.

An $n$-qubit diagonal unitary matrix can be represented as $\sum\limits_{i = 0}^{2^n - 1} {\lambda_i|i\rangle \langle i|}$ where $\lambda_i^,s$ are the eigenvalues of the matrix and $|\lambda_i|=1$.

Since the diagonal elements of a diagonal matrix are its eigenvalues, the diagonal elements of a diagonal Hermitian quantum gate are either +1 or -1.
\subsection{The Properties of $\mathrm{C^{k}Z}$ Gates}
\label{sec:hermiti}
\begin{figure}
\centering
\[
\Qcircuit @C=1.0em @R=1.4em{
& \ctrl{1}&\qw\\
& \ctrl{0}&\qw
}
\]
\caption{The circuit representation of CZ gate.}
\label{fig:cz}
\end{figure}
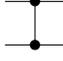

Since $\mathrm{C^kZ}$ gates are used in the proposed decompostion method, some of their properties are discussed in this section.
The matrix representation of the CZ gate is as follows:
\[
\text{CZ}=%
\begin{bmatrix}
1 & 0& 0&0\\
0 & 1&0&0\\
0&0&1&0\\
0&0&0&-1
\end{bmatrix}.\]
Figure~\ref{fig:cz} shows the circuit representation of CZ gate.

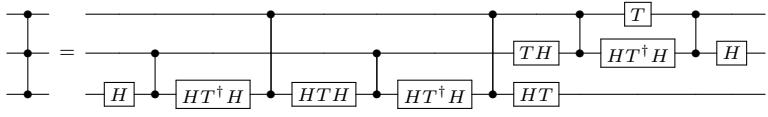
\begin{figure*}
\scriptsize
\centering
\[
\Qcircuit @C=1.0em @R=0.7em {
&\ctrl{1} & \qw & & & \qw & \qw &\qw &\ctrl{2} &\qw &\qw &\qw &\ctrl{2} &\qw &\ctrl{1} &\gate{T} &\ctrl{1} &\qw &\qw \\
&\ctrl{1} & \qw &=& & \qw & \ctrl{1} & \qw & \qw &\qw &\ctrl{1} &\qw &\qw &\gate{TH} & \ctrl{0} & \gate{HT^\dag H}&\ctrl{0} & \gate{H}&\qw\\
&\ctrl{0} & \qw & & &\gate{H} &\ctrl{0}&\gate{HT^\dag H} &\ctrl{-2}&\gate{HTH}&\ctrl{-1}&\gate{HT^\dag H}&\ctrl{-2}&\gate{HT}&\qw &\qw &\qw &\qw &\qw
}
\]
\caption{The decomposition of $\mathrm{C^2Z}$ gate to CZ and single-qubit gates~\cite{Shende09}.}
\label{fig:C2Z}
\end{figure*}

In this paper, $Z$ gate is also denoted as $\mathrm{C^{0}Z}$ gate. The $\mathrm{C^{k}Z}$ gates for $k\ge 0$ are symmetric with respect to exchanging qubits. They negate the input qubits when all of them are in the state $|1\rangle$ and otherwise leave them unchanged. These gates are diagonal and self-inverse and they all commute with each other. Consider a circuit $\mathcal{C}$ that solely consists of these gates. It is worth mentioning that if any of these gates may occur at most once in $\mathcal{C}$, as these gates commute and if one gate appears more than once in $\mathcal{C}$, they can be moved next to each other and be canceled out. The $\mathrm{C^{k}Z}$ gates for $k>1$ can in turn be decomposed to CZ and single qubit gates. As an example, the decomposition of $\mathrm{C^2Z}$ gate to CZ and single-qubit gates by the approach of~\cite{Shende09} is shown in Figure~\ref{fig:C2Z}.

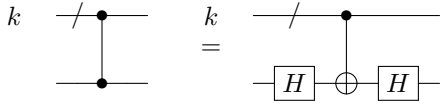
\begin{figure}
\centering
\[
\Qcircuit @C=0.8em @R=0.0000001em @!R{
&k&&&/\qw	&\ctrl{2}	&\qw   	&  \qw							&&&k&&		&/\qw				 &\ctrl{2}	&\qw		 			&\qw		\\
&	&&&	&       	&  &	&			 			&&=	&&	&&&		&        			 &          	 &         				&		 \\
&	&&&\qw	&\ctrl{0} 	&\qw &\qw 	&  		&&	& &	&\gate{H}	&\targ	&\gate{H}	&\qw		
}
\]
\caption{Circuit equivalence of $\mathrm{C^kZ}$ gates and $\mathrm{C^kNOT}$ gates.}
\label{fig:czcnot}
\end{figure}

Using the equation $HXH=Z$, a $\mathrm{C^kZ}$ gate can be easily replaced by a $\mathrm{C^kNOT}$ gate in the middle and two Hadamard gates which act on target qubit, as shown in Figure~\ref{fig:czcnot}.
Similarly, $\mathrm{C^kNOT}$ gates can be replaced by $\mathrm{C^kZ}$ gates at the cost of inserting two Hadamard gates.

\section{Proposed approach}
\label{sec:proposed}
In this section, some definitions and notations are introduced before proceeding to the main part of the paper.
\subsection{Definitions and Notations}
\label{sec:def}
Consider $\mathbb{D}_n=\{d_n, n \in N\}$ where each $d_n$ is a diagonal Hermitian matrix on $n$ qubits.
For a given $d_n$, its $i^{th}$ diagonal element is denoted as $\lambda_i$ for $0\leq i \leq 2^{n}-1$. Based on the definition of $d_n$, each element of it is either +1 or -1. Without the loss of generality, it can be assumed that the first element of a given $d_n$, i.e., $\lambda_0=1$. If $\lambda_0=-1$, one can substitute $d_n$ with $-d_n$.

This new matrix is equivalent to the old one up to a global phase. As a result, $2^{n}-1$ independent diagonal elements of $d_n$, i.e., $\lambda_1,...,\lambda_{2^n-1}$, are left.

The following two functions will be used later in the proposed approach.
\begin{align*}
b=\text{Binary}(d_n)
\,,d_n=\text{DiagMat}(b)
\end{align*}
The first function, takes a matrix ($d_n$) and returns a corresponding $(2^n-1)$-bit binary number $b$. To this end, the elements with the +1 and -1 values of $d_n$ are first replaced by 0 and 1 respectively and then the sequence of the diagonal elements of this matrix (excluding $\lambda_0$) are written in $b$ from right to left. For example, $b=(1, 0, 0)=\text{Binary(CZ)}$. We call $b$ the binary representation of $d_n$. The second function, $d_n=\text{DiagMat}(b)$ is the reverse of the previous one which takes a binary number $b$ and returns the corresponding matrix. For example, CZ=DiagMat(1, 0, 0).

When a number of $d_n$ gates are cascaded, a matrix is produced whose $i^{th}$ diagonal element is obtained by the multiplication of all of the $i^{th}$ diagonal elements of these gates. It can be readily verified that the multiplication operation on the set $\{  - 1,1\}$ is equivalent to XOR operation on the set $\{ 1,0\}$
where algebric value -1 corresponds to binary value 1 and algebraic value 1 corresponds to binary value 0.

Consider $z=\mathrm{XOR}_{1\leq i \leq k}(x^i)$ a function which takes $k$ $n$-bit binary numbers, $x^i$, $1\leq i \leq k$ and returns a binary number $z$ where $z_j$, i.e., the $j^{th}$ bit of $z$, $0\leq j\leq n-1$ is computed by xoring all of the $j^{th}$ bits of $x^i$.
Therefore, in order to compute the multiplication of $d_n^i$ gates, $1\leq i \leq k$, the following equation is used:

\begin{equation}
\prod\nolimits_{i = 1}^k {d_n^i }  = \text{DiagMat(XOR}_{1 \le i \le k} (\text{Binary}(d_n^i)))
\label{eq:eq1}
\end{equation}

 The number of different $\mathrm{C^kZ}$ gates for $0 \leq k \leq n-1$ which act on $n$ qubits is equal to $2^{n}-1$.
 To prove this, consider that there are $\binom{n}{1}$ possible $\mathrm{C^0Z}$ gates, $\binom{n}{2}$ possible $\mathrm{C^1Z}$ gates, $\binom{n}{3}$ possible $\mathrm{C^{2}Z}$ gates, ..., and $\binom{n}{n}$ possible $\mathrm{C^{n-1}Z}$ gates in the same way. On the other hand, the number of $k$-combinations for all $k$ equals to the number of subsets of a set with $n$ members:
\[\binom{n}{0}+\binom{n}{1}+\binom{n}{2}+...+\binom{n}{n}=2^{n}\]

Therefore, the number of the different $\mathrm{C^kZ}$ gates is equal to:
\begin{equation}
\binom{n}{1}+\binom{n}{2}+...+\binom{n}{n}=2^{n}-1.\nonumber
\label{eq:num}
\end{equation}

In the rest of the paper, to specify different $\mathrm{C^kZ}$ gates which act on $n$ qubits, they are indexed.
To this end, first the qubits in an $n$-qubit circuit are numbered from 0 to $n$-1 from top to bottom and for each $\mathrm{C^kZ}$ gate a corresponding $n$-bit binary string, denoted by $\mathbf{b}$ can be produced. The $j^{th}$ bit of $\mathbf{b}$ from right, $0\leq j \leq n-1$ is 1, if $\mathrm{C^kZ}$ is applied on the $j^{th}$ qubit of the circuit and is 0, otherwise. Then, the $\mathrm{C^kZ}$ gate is shown by $i$, the decimal value of $\mathbf{b}$, which ranges from 1 to $2^n-1$, as $\mathbf{CZ}_i$. For example, in a two-qubit circuit, the binary strings of the gates $Z\otimes I$, $I\otimes Z$, and CZ are produced as $01$, $10$ and $11$ and they are denoted as $\mathbf{CZ}_1$, $\mathbf{CZ}_2$ and $\mathbf{CZ}_3$, respectively.

\subsection{\label{sec:decom}Decomposition of $\mathbb{D}_n$ Gates}
In this section, we propose an approach to decompose a given $d_n$ matrix to $\mathbf{CZ}_i$ gates for $1\leq i\leq 2^n-1$.
\begin{lemma}
\label{lemma1}
The set of $\mathrm{{C^{k}Z}}$ gates for $k\geq0$ are independent, i.e., none of them can be decomposed to a combination of the other $\mathrm{{C^{k}Z}}$ gates.
\end{lemma}
\begin{proofs}
The proof is done by contradiction. Suppose a gate $\mathrm{C^{k}Z}$ can be decomposed to a set of gates that may consist of $\mathrm{C^{0}Z, C^1Z, ..., C^{k-1}Z}$ gates. This set of gates produces a circuit which is called $\mathcal{C}$. Now, consider $\mathrm{C^{m}Z}$ a gate which has the smallest number of control qubits (i.e., $m$) among all these gates.
 Let the qubits on which $\mathrm{C^{m}Z}$ gate acts be $|1\rangle$ and the remaining input qubits be $|0\rangle$. As other gates in $\mathcal{C}$ have more control lines than $\mathrm{C^{m}Z}$ gate and at least one of them is in the state $|0\rangle$, none of them activates. As $\mathrm{C^{m}Z}$ gate acts, the output of $\mathcal{C}$ is the negative of its inputs. On the other hand, $\mathrm{C^{k}Z}$ gate has no effect on its input because at least one of its control lines is in the $|0\rangle$ state. Therefore, there is at least one input state for which the outputs of $\mathcal{C}$ and $\mathrm{C^{k}Z}$ gate differ which is a contradiction. Hence, $\mathrm{C^{k}Z}$ gates for $k\geq0$ are independent.
\end{proofs}
\begin{theorem}
Every $d_n$ gate can be decomposed to a combination of $\mathbf{CZ}_i$ gates for $1 \leq i \leq 2^n-1$.
\end{theorem}
\begin{proofs}
The Binary function as introduced in Section~\ref{sec:def} returns a $(2^n-1)$-bit binary number. On the other hand,  there are $2^n-1$ independent $\mathbf{CZ}_i$ gates, as shown in Lemma~\ref{lemma1}. Therefore, the binary representations of these $2^n-1$ gates are also independent and form a basis for the vector space produced by $(2^n-1)$-bit binary numbers when their combination is performed by the XOR operator. Therefore, each diagonal Hermitian quantum gate can be written as a combination of $\mathbf{CZ}_i$ gates where their multiplication is performed using Equation~$\left(\ref{eq:eq1}\right)$. These gates may have a coefficient of 1 if they exist and a coefficient of 0 if they do not exist in the decomposed circuit. In other words, the following equation should be used:
\begin{equation}
\text{Binary}(d_n)=
\text{XOR}_{1 \le i \le 2^n-1 }(a_i.\text{Binary}(\textbf{CZ}_i))
\label{eq:eqset}
\end{equation}

where ${a_i}^,s$ are binary variables and dot represents the logical AND operator which is applied on $a_i$ and each bit of the binary representation of $\textbf{CZ}_i$. It is worth mentioning that as we had assumed that the first elements of the diagonal Hermitian matrices are always +1 and the first element of all $\mathbf{CZ}_i$ gates are +1, the equation corresponding to them always satisfies.

By solving Equation~(\ref{eq:eqset}), $a_i^,s$ can be found.
\end{proofs}

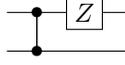
\begin{figure}
\centering
\[
\Qcircuit @C=1.0em @R=.7em {
&\ctrl{1} &\gate{Z}&\qw \\
& \ctrl{-1} &\qw & \qw
}
\]
\caption{The circuit that implements matrix $A=|0\rangle \langle 0|-|1\rangle \langle 1|-|2\rangle \langle 2|+|3\rangle \langle 3|$.}
\label{fig:fig2}
\end{figure}

Example 1 clarifies the discussion.\\
\textbf{Example 1.} Suppose a two-qubit diagonal Hermitian matrix $A=|0\rangle \langle 0|+|1\rangle \langle 1|-|2\rangle \langle 2|+|3\rangle \langle 3|$ is given. The function Binary(A) returns (0,1,0). In order to apply the proposed approach, the gates $\mathbf{CZ}_1$, $\mathbf{CZ}_2$ and $\mathbf{CZ}_3$ are considered. (1, 1, 0)=Binary($\mathbf{CZ}_1$), (1, 0, 1)=Binary($\mathbf{CZ}_2$) and (1, 0, 0)=Binary($\mathbf{CZ}_3$). We have the following set of equations:
\begin{equation}
a_{1}.(1, 1, 0)\oplus a_{2}.(1, 0, 1) \oplus a_{3}.(1, 0, 0)=(0, 1, 0)\nonumber
\end{equation}
where $a_i^,s$ are binary variables.
Therefore, the following equation set must be solved:
\begin{equation}
\left\{{\begin{array}{*{20}c}
   {a_1.(1)\oplus a_2.(1) \oplus a_3.(1) =  0}\\
   {a_1.(1)\oplus a_2.(0) \oplus a_3.(0) =  1}\\
   {a_1.(0)\oplus a_2.(1) \oplus a_3.(0) =  0}\\
\end{array}}\right.
\label{eq:set}
\end{equation}
By solving Equation~(\ref{eq:set}), $a_1$, $a_2$ and $a_3$ are found as 1, 0 and 1, respectively. This solution set implies that the gates $\mathbf{CZ}_1$ and $\mathbf{CZ}_3$ exist and $\mathbf{CZ}_2$ does not exist in the synthesized circuit, respectively. The circuit that implements the $A$ gate is shown in Figure~\ref{fig:fig2}.

\subsection{Cost Analysis}
\label{sec:cost}
To decompose a given $d_n$ gate, the proposed approach in the worst case produces a circuit that consists of all possible $2^n-1$ $\mathbf{CZ}_i$ gates (as explained in Section~\ref{sec:def}). Without using any ancilla qubits, $\mathrm{C^{n-1}Z}$ gates can be decomposed to $O(n^2)$ elementary gates based on the results of~\cite[Corollary 7.6]{Barenco95}. Therefore, the cost complexity of the proposed approach in the worst case is $O(n^22^n)$ in terms of elementary gates.

On the other hand, as mentioned in Section~\ref{sec:ckz}, the $\mathrm{C^kZ}$ gates can be replaced by $\mathrm{C^kNOT}$ gates at the cost of inserting two Hadamard gates~\cite{Shende09} and there are many studies on the efficient decomposition of $\mathrm{C^kNOT}$ gates in the literature. For example, in~\cite{Barenco95} it is shown that in a circuit on $n$ qubits, using one ancilla, a $\mathrm{C^{n-2}NOT}$ gate, $n\ge7$ can be decomposed to $8(n-2)$ Toffoli gates and using $m-2$ ancilla qubits, $m \in \{3,4,..,\lceil n/2 \rceil\}$, $n\ge5$, a $\mathrm{C^mNOT}$ gate  can be decomposed to $4(m-2)$ Toffoli gates. These studies can be applied to decrease the worst case cost complexity of the proposed approach.

\section{Experimental Results} \label{sec:results}
As there are $2^n-1$ independent diagonal elements in a $\mathbb{D}_n$ matrix, each of which can take two distinct values, the number of all possible $\mathbb{D}_n$ gates is $2^{2^n-1}$. For example, the numbers of all possible $\mathbb{D}_2$, $\mathbb{D}_3$ and $\mathbb{D}_4$ gates are 8, 128 and 32678, respectively.

The proposed approach and the approach of~\cite{bullock2004} for decomposing $\mathbb{D}_i$ gates for $i=2,3,4$ were implemented in MATLAB and MAPLE, respectively on a workstation with 4GB RAM and Core i5 2.40GHz CPU.
Table~\ref{table} compares the results of the proposed approach and the approach of~\cite{bullock2004} to decompose some members of $\mathbb{D}_i$ for $i=2, 3, 4$.

\begin{figure}
\centering
\scriptsize
\[
\Qcircuit @C=0.7em @R=0.5em @!R {
&\qw \gategroup{7}{10}{8}{14}{.9em}{--}	& \qw  		&\qw 			&\qw		&\qw 			&\qw 		&\qw 			&\ctrl{3} 	&\qw 			& \qw  		 &\qw 			&\qw		&\qw 			&\qw 		&\qw 			&\ctrl{3}		&\qw\\
&\qw 					& \qw 		&\qw 			&\ctrl{2}	&\qw 			&\qw 		&\qw			&\qw 		&\qw 			& \qw 		&\qw 			 &\ctrl{2}	&\qw 			&\qw 		&\qw			&\qw 			&\qw\\
&\qw					&\ctrl{1}	&\qw 			&\qw 		&\qw			&\ctrl{1}	&\qw 			&\qw 		&\qw			&\ctrl{1}	&\qw 			 &\qw 		&\qw			&\ctrl{1}	&\qw 			&\qw 	 		&\qw\\
&\gate{R_z(o)H}				&\ctrl{0} 	&\gate{H R_z(n)H} 	&\ctrl{0}	&\gate{H R_z(m) H} 	&\ctrl{0}	&\gate{H R_z(l) H}	&\ctrl{0}	&\gate{HR_z(k)H}	 &\ctrl{0} 	&\gate{H R_z(j)H} 	&\ctrl{0}	&\gate{H R_z(i) H} 	&\ctrl{0}	&\gate{H R_z(h) H}	&\ctrl{0}		&\qw\\
&					&		&			&		&			&		&			&		&			&		&			&		&			&		&			 &		&	&\\
&					&		&			&		&			&		&			&		&			&		&			&		&			&		&			 &		&	&\\
&\qw \gategroup{6}{1}{9}{15}{2.3em}{..}	& \qw  		&\qw 			&\ctrl{2} 	&\qw 			&\qw 		&\qw 			&\ctrl{2} 	&\qw 			 &\ctrl{1} 	&\qw 			&\ctrl{1}	&\gate{R_z(a)}		&\qw		&			&		&	&\\
&\qw 					& \ctrl{1} 	&\qw 			&\qw 		&\qw 			&\ctrl{1} 	&\qw			&\qw 		&\gate{R_z(c)H} 	& \ctrl{0} 	&\gate{H R_z(b)H}	&\ctrl{0} 	&\gate{H}		&\qw		&			&		&	&\\
&\gate{R_z(g)H}				&\ctrl{0} 	&\gate{H R_z(f)H} 	&\ctrl{0}	&\gate{H R_z(e) H} 	&\ctrl{0}	&\gate{H R_z(d) H}	&\ctrl{0}	&\gate{H}		&\qw		 &\qw 			&\qw 		&\qw 			&\qw		&			&		&	&\\
&\gate{H}				&\qw 		&\qw 			&\qw 		&\qw			&\qw		&\qw 			&\qw 		&\qw 			&\qw		&\qw			 &\qw 		&\qw 			&\qw		&			& 		&	&
}
\]
\caption{The general circuit structure to implement four-qubit diagonal gates by~\cite{bullock2004} using CZ and single-qubit gates, up to a global phase. The dashed and the dotted parts of the figure can be assumed as general circuit structures for two and three-qubit diagonal gates, respectively. The circuit of the bottom is the continuation of the top circuit.}
\label{fig:previous}
\end{figure}
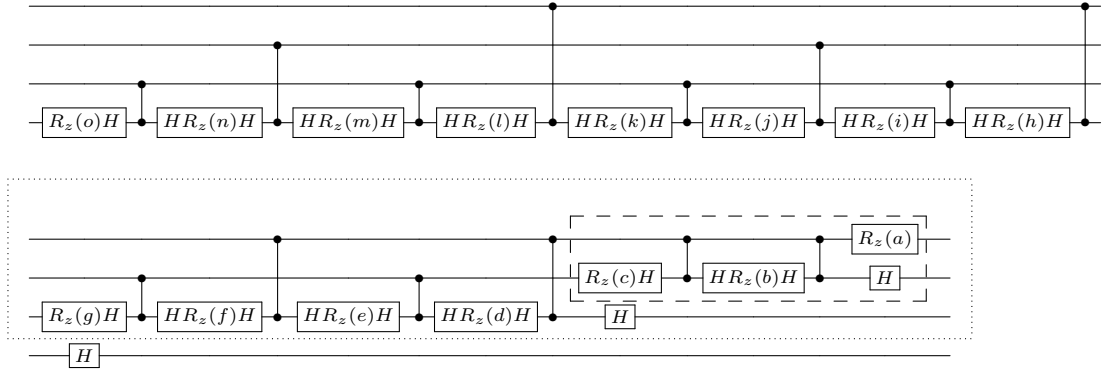

 In this table, each matrix is denoted by a number equal to the decimal value of its binary representation. For example, $(1,0,0)$=CZ is denoted by 4. The second column of the table contains the indices of the produced basis state in the proposed approach (i.e., $\mathbf{CZ}_i$ gates), as obtained by the procedure explained in Section~\ref{sec:proposed}. The angles of the produced $R_z(\alpha)$ gates for the approach of~\cite{bullock2004}, according to the circuit structure of Figure~\ref{fig:previous}, are determined in the third column.

The numbers of required CZ and single-qubit gates in terms of rotation gates around $y$ and $x$ axes for each circuit are reported and compared for these two approaches. Possible optimizations are performed on the circuits produced by the approach of~\cite{bullock2004}. The optimizations of the CNOT gates for this approach are performed by cancelling CNOT gates whenever possible using the transformation rules of these gates~\cite{iwama2002transformation}. The utilized transformation rules are cancelling two adjacent identical CNOT gates and changing the order of independent CNOT gates. Moreover, by using the equation of $H^2=I$, whenever two adjacent Hadamard gates appear in the circuit, they cancel and therefore the number of single-qubit gates can be decreased.

In this paper, the library of CZ and rotation gates around $y$ and $x$ axes is considered. To decompose the circuits based on the previously known elementary library which includes CNOT gate, CZ gates can be readily replaced by a middle CNOT gate and two Hadaramd gates which act on the target qubit (as shown in Figure~\ref{fig:czcnot}). After this replacement, the circuits should be further optimized by cancelling two adjacent Hadamard gates.

The experimental results on all possible $\mathbb{D}_2$, $\mathbb{D}_3$ and $\mathbb{D}_4$ gates show that the proposed approach improves the cost of CZ and single-qubit gates of the method of~\cite{bullock2004} by $37.4\%$, $34.3\%$, $10.9\%$, $31\%$, $4.6\%$ and $24.7\%$, respectively on average. As can be seen, when the number of qubits increases, the average improvement decreases. However, as Table~\ref{table} shows, with the increase of the number of qubits, there are always some cases where the proposed approach leads to significantly better results.

\begin{center}
\scriptsize
\begin{longtable}{|l||c|l|c|c|l|c|c|c|c|}%
\caption{The comparison of obtained circuits by the proposed approach and the method of [Bullock and Markov 2004] to decompose some members of $\mathbb{D}_2$, $\mathbb{D}_3$ and $\mathbb{D}_4$ gates. $a,b,...,0$ are the angles of $R_z(\alpha)$ gates of Figure~\ref{fig:previous}. The single qubit gates are of types $R_x$ and $R_y$.}\label{table}\\

\hline\multirow{2}{*}{$n$} 	&\multirow{2}{*}{Matrix}	&\multicolumn{3}{c|}{The Proposed Approach}	&\multicolumn{3}{c|}{\cite{bullock2004}}	 &\multicolumn{2}{c|}{Imp. (\%)}						\\	
			& 				&Basis		&\#CZ		&\#1qu	&$a,b,c$						&\#CZ		&\#1qu	&\#CZ		&\#1qu\\
\hline\hline\endfirsthead

\multicolumn{10}{c}%
{{\bfseries \tablename\ \thetable{} -- continued from previous page}} \\
\hline\multirow{2}{*}{$n$} 	&\multirow{2}{*}{Matrix}	&\multicolumn{3}{c|}{The Proposed Approach}	&\multicolumn{3}{c|}{\cite{bullock2004}}	 &\multicolumn{2}{c|}{Imp. (\%)}							\\	
			& 				&Basis		&\#CZ		&\#1qu	&$a,b,c,d,e,f,g,h,i,j,k,l,m,n,o$						&\#CZ		&\#1qu	&\#CZ		&\#1qu\\
\hline\hline\endhead
\hline\hline\multicolumn{10}{|c|}{Continued on next page} \\
\hline\endfoot
\hline\endlastfoot
	
2	&1		&2,3		&1		&3		&$\frac{-\pi}{2},	\frac{\pi}{2},		\frac{\pi}{2}$						&2		&12	&50		&75\\
	&2		&1,3		&1		&3		&$\frac{\pi}{2},	\frac{\pi}{2},		\frac{-\pi}{2}$						&2		&12	&50		&75\\
	&3		&1,2		&0		&6		&$0,			\pi,			0$							&2		&9	&100		&33.3\\
	&4		&3		&1		&0		&$\frac{\pi}{2},	\frac{-\pi}{2},		\frac{\pi}{2}$						&2		&12	&50		&100\\
	&7		&1,2,3		&1		&6		&$\frac{\pi}{2},	\frac{\pi}{2},		\frac{\pi}{2}$						&2		&12	&50		&50\\\hline\hline
\hline\hline		
\multirow{2}{*}{$n$} 	&\multirow{2}{*}{Matrix}	&\multicolumn{3}{c|}{The Proposed Approach}	&\multicolumn{3}{c|}{\cite{bullock2004}}	 &\multicolumn{2}{c|}{Imp. (\%)}						\\	
			& 				&Basis		&\#CZ		&\#1qu	&$a,b,c,d,e,f,g$						&\#CZ		&\#1qu	&\#CZ		 &\#1qu\\\hline\hline
3	&15		&1,2,4,6	&1		&9		&$\frac{-\pi}{2},	\frac{\pi}{2},	0,		\frac{\pi}{2},	\frac{\pi}{2},	0,	 	0		$	&6		&24		&83.3	 &62.5						\\
	&18		&2,3,5,6	&3		&3		&$0,			\frac{\pi}{2},	0,		\frac{-\pi}{2},	0,		\frac{\pi}{2},	0		$	&6		&21		&50	&85.7						 \\
	&20		&5,6		&2		&0		&$0,			\frac{\pi}{2},	0,		0,		\frac{-\pi}{2},	0,		\frac{\pi}{2}	$	&6		&18		&66.6	&100						 \\
	&27		&1,2,4,5	&1		&9		&$0,			\frac{\pi}{2},	\frac{-\pi}{2},	0,		\frac{\pi}{2},	\frac{\pi}{2},	0		$	&6		&21		 &83.3	&57.1						\\
	&45		&1,4		&0		&6		&$0,			0,		0,		\pi,		0,		0,	 	0		$	&2		&9		&100	&33.3						 \\		
	&51		&2,4		&0		&6		&$0,			0,		0,		0,		0,		\pi,	 	0		$	&2		&9		&100	&33.3									 \\
	&54		&2,5		&1		&3		&$0,			\frac{\pi}{2},	\frac{\pi}{2}, 	0,		\frac{-\pi}{2},	\frac{\pi}{2}, 	0		$	&6		&21		 &83.3	&85.7									\\
	&65		&4,5,6		&2		&3		&$0,			\frac{-\pi}{2},	0,		0,		\frac{\pi}{2},	0,		\frac{\pi}{2}	$	&6		&18		&66.6	 &83.3									 \\
	&99		&2,4,5		&1		&6		&$0,			\frac{-\pi}{2},	\frac{\pi}{2},	0,		\frac{\pi}{2},	\frac{\pi}{2},	0		$	&6		&21		 &83.3	&71.4									 \\
	&113		&3,4,6		&2		&3		&$\frac{\pi}{2},	\frac{-\pi}{2},	0,		0,		0,		\frac{\pi}{2},	\frac{\pi}{2}	$	&4		&21		&50	 &85.7							\\
	\hline\hline 																				
\multirow{2}{*}{$n$} 	&\multirow{2}{*}{Matrix}	&\multicolumn{3}{c|}{The Proposed Approach}	&\multicolumn{3}{c|}{\cite{bullock2004}}	 &\multicolumn{2}{c|}{Imp. (\%)}						\\	
			& 				&Basis 		&\#CZ		&\#1qu	&$a,b,c,d,e,f,g,h,i,j,k,l,m,n,o$						&\#CZ		&\#1qu	&\#CZ		 &\#1qu\\\hline\hline
4	&4680		&3,6,9,12	&4		&0		&$	0,	0,	0,	\frac{\pi}{2},	0,	0,	0,		0,	0,	\frac{-\pi}{2},	0,	\frac{\pi}{2},	0,	0,	0														 $&8			&15		&50		&100		\\
	&10376		&11,12		&7		&27		&$	0,	0,	0,	\frac{\pi}{4},	\frac{-\pi}{4},	\frac{\pi}{4},	\frac{\pi}{4},		0,	\frac{-\pi}{4},	 \frac{\pi}{4},	0,	0,	\frac{-\pi}{4},	\frac{-\pi}{4},	\frac{\pi}{2}								$&10			&33		&30		&18.1		\\
	&14602		&1,5,8,10	&2		&6		&$	\frac{\pi}{4},	\frac{-\pi}{4},	\frac{-\pi}{4},	\frac{\pi}{4},	\frac{-\pi}{4},	\frac{\pi}{4},	\frac{-\pi}{4},		 \frac{\pi}{4},	\frac{\pi}{4},	\frac{\pi}{4},	\frac{\pi}{4},	\frac{\pi}{4},	\frac{-\pi}{4},	\frac{-\pi}{4},	\frac{\pi}{4}		$&14			&54		 &85.7		&88.8		\\
	&21760		&1,9		&1		&3		&$	\frac{\pi}{2},	0,	0,	0,	0,	0,	0,		\frac{\pi}{2},	0,	0,	0,	0,	0,	0,	\frac{-\pi}{2}														 $&2			&9		&50		&66.6		\\
	&23280		&2,9		&1		&3		&$	0,	\frac{\pi}{2},	\frac{\pi}{2},	0,	0,	0,	0,		0,	0,	0,	\frac{-\pi}{2},	\frac{\pi}{2},	0,	0,	0													 $&6			&18		&83.3		&83.3		\\
	&24428		&1,4,5,10	&2		&6		&$	\frac{\pi}{4},	\frac{\pi}{4},	\frac{-\pi}{4},	\frac{\pi}{4},	\frac{\pi}{4},	\frac{\pi}{4},	\frac{\pi}{4},		 \frac{\pi}{4},	\frac{\pi}{4},	\frac{-\pi}{4},	\frac{-\pi}{4},	\frac{\pi}{4},	\frac{-\pi}{4},	\frac{\pi}{4},	\frac{-\pi}{4}		$&14			&54		 &85.7		&88.8		\\
	&27030		&1,2,4,8	&0		&12		&$	0,	0,	0,	0,	0,	0,	0,		0,	0,	0,	0,	\pi,	0,	0,	0																	 $&6			&6		&100		&-100		\\
	&38460		&2,4,9		&1		&6		&$	0,	0,	0,	0,	\frac{\pi}{2},	\frac{\pi}{2},	0,		0,	0,	\frac{-\pi}{2},	0,	0,	\frac{\pi}{2},	0,	0													 $&10			&24		&90		&75		\\
	&40044		&3,4,10		&2		&3		&$	0,	0,	0,	0,	0,	\frac{\pi}{2},	\frac{\pi}{2},		0,	\frac{\pi}{2},	\frac{-\pi}{2},	0,	0,	0,	0,	0													 $&8			&40		&75		&92.5		\\
	&43520		&9		&1		&0		&$	\frac{\pi}{2},	0,	0,	0,	0,	0,	0,		\frac{-\pi}{2},	0,	0,	0,	0,	0,	0,	\frac{\pi}{2}														 $&2			&9		&50		&100		\\
	&49258		&9,6,8		&2		&3		&$	\frac{-\pi}{4},	\frac{-\pi}{4},	\frac{\pi}{4},	\frac{-\pi}{4},	\frac{\pi}{4},	\frac{-\pi}{4},	\frac{\pi}{4},		 \frac{\pi}{4},	\frac{\pi}{4},	\frac{-\pi}{4},	\frac{\pi}{4},	\frac{\pi}{4},	\frac{-\pi}{4},	\frac{\pi}{4},	\frac{\pi}{4}		$&14			&54		 &85.7		&94.4		\\
	&51884		&4,5,6,9,10	&4		&3		&$	0,	0,	0,	0,	\frac{\pi}{2},	0,	\frac{\pi}{2},		0,	0,	0,	\frac{-\pi}{2},	0,	0,	0,	\frac{\pi}{2}													 $&8			&15		&50		&80		\\
	&63120		&2,5,6,9,10	&4		&3		&$	\frac{\pi}{2},	0,	\frac{\pi}{2},	0,	0,	0,	0,		0,	\frac{-\pi}{2},	0,	0,	0,	\frac{\pi}{2},	0,	0													 $&6			&18		&33.3		&83.3		\\
	&63916		&1,4,6,9,10	&3		&6		&$	\frac{\pi}{4},	\frac{-\pi}{4},	\frac{\pi}{4},	\frac{\pi}{4},	\frac{\pi}{4},	\frac{\pi}{4},	\frac{\pi}{4},		 \frac{\pi}{4},	\frac{\pi}{4},	\frac{\pi}{4},	\frac{-\pi}{4},	\frac{-\pi}{4},	\frac{-\pi}{4},	\frac{-\pi}{4},	\frac{\pi}{4}		$&14			&54		 &78.5		&88.8		\\
	&64598		&2,4,6,8,9	&2		&9		&$	\frac{\pi}{4},	\frac{-\pi}{4},	\frac{\pi}{4},	\frac{-\pi}{4},	\frac{-\pi}{4},	\frac{\pi}{4},	\frac{\pi}{4},		 \frac{-\pi}{4},	\frac{\pi}{4},	\frac{\pi}{4},	\frac{\pi}{4},	\frac{\pi}{4},	\frac{\pi}{4},	\frac{\pi}{4},	\frac{-\pi}{4}		$&14			&54		 &85.7		&83.3		\\
\end{longtable}																								
\end{center}																						
\section{Conclusions}
\label{sec:conc}
In this paper, we presented a new approach to decompose $n$-qubit diagonal Hermitian gates, denoted by $\mathbb{D}_n$. First, a function was defined that takes the diagonal elements of a specific $n$-qubit diagonal Hermitian quantum gate, $d_n$, and returns a binary representation for it. Then we showed that $\mathbb{D}_n$ gates can be decomposed to a circuit that is solely constructed of $\mathrm{C^kZ}$ gates for $0\leq k\leq n-1$. Moreover, it was proved that the binary representations of this set of gates, form a basis for the vector space that is produced by the binary representations of $\mathbb{D}_n$ gates, denoted by $b$, where their combination is performed using logical XOR gates. After that, it was shown that the decomposition problem of $\mathbb{D}_n$ gates can be mapped to writing $b$ in that basis where the coefficients of 1 and 0 imply that the corresponding basis state exists and does not exists in the produced circuit, respectively. Experimental results on all possible two, three and four-qubit diagonal Hermitian gates show that the proposed approach improves the cost of CZ and single-qubit gates of best previous method for synthesizing arbitrary diagonal gates by $37.4\%$, $34.3\%$, $10.9\%$, $31\%$, $4.6\%$ and $24.7\%$, respectively on average. As a future work, applying some post processing after the proposed decomposition to optimize the produced circuits is being considered. Moreover, we are seeking more general unitary matrices to which the proposed perspective can be applied.

\footnotesize

\end{document}